# Activity of comets: Gas Transport in the Near-Surface Porous Layers of a Cometary Nucleus

Yuri V. Skorov*[a,b], Rik van Lieshout[c], Jürgen Blum[b], Horst U. Keller[b]

[a] – Max Planck Institute for Solar System Research, Max-Planck-Str. 2, D-37191, Katlenburg-Lindau, Germany
[*] Corresponding Author. E-mail address: skorov@mps.mpg.de
[b] –Institute for Geophysics and Extraterrestrial Physics,
Technical University of Braunschweig,
Mendelssohn-Str. 3, D-38106 Braunschweig, Germany
[c] – Astronomical Institute Anton Pannekoek, University of Amsterdam, Science Park 904, 1098 XH Amsterdam, The Netherlands

**Abstract**

The gas transport through non-volatile random porous media is investigated numerically. We extend our previous research of the transport of molecules inside the uppermost layer of a cometary surface (Skorov and Rickmann, 1995; Skorov et al. 2001). We assess the validity of the simplified capillary model and its assumptions to simulate the gas flux trough the porous dust mantle as it has been applied in cometary physics. A new microphysical computational model for molecular transport in random porous media formed by packed spheres is presented. The main transport characteristics such as the mean free path distribution and the permeability are calculated for a wide range of model parameters and compared with those obtained by more idealized models. The focus in this comparison is on limitations inherent in the capillary model. Finally a practical way is suggested to adjust the algebraic Clausing formula taking into consideration the nonlinear dependence of permeability on layer porosity. The retrieved dependence allows us to accurately calculate the permeability of layers whose thickness and porosity vary in the range of values expected for the near-surface regions of a cometary nucleus.

**Introduction**

Comets are fabulous eye-catching members of the Solar System because of their gas activity when they come close to the sun. One of the most demanding and pressing problems of cometary physics is the question how their activity works in detail. Sublimation of volatiles (ices), that constitute a substantial part of the cometary nucleus, under the influence of solar radiation triggers the simultaneous release of non-volatile components and leads to the formation of the gas-dust coma, the size of which is thousands of times larger than the nucleus itself.

Coma observations have provided almost the only source of our knowledge of comets. This only changed with the advent of space mission. Starting with the flybys of comet 1P/Halley in 1986 (Keller et al., 1987) a few snapshots of the inner coma and nucleus were delivered by spacecraft. Consequently, the focus of cometary research has traditionally concentrated on the study of processes occurring in the coma. Numerous photometric, spectroscopic, and polarimetric observations have allowed us to develop sophisticated and detailed theoretical models describing the physical and chemical processes in a coma as well as its large-scale spatial dynamic structure forming the plasma and dust tails. It is important to note that all these models of the coma in one form or another are based on the results of models that determine the gas activity of the nucleus. These latter models can be very simple. For example, the gas production rate of the cometary nucleus, consisting of water ice can be calculated from the average energy balance at the surface. In the simplest case the thermal conductivity of core material and its thermal re-radiation are often not taken into account. Thus all the absorbed solar energy is used for sublimation of water ice, and the rate of ice sublimation $E_i(T)$ is calculated by the known Hertz-Knudsen equation

$$E_i(T) = p_{sat}(T)\sqrt{\frac{m}{2\pi k_B T}} L(T) , \qquad (1)$$

where $k_B$ is the Boltzmann constant, $m$ is water molecule mass, $L(T)$ is evaporation heat, $p_{sat}$ is the equilibrium vapor pressure at the surface temperature $T$.

These model assumptions provide the basis for the "conglomerate" icy model of a homogeneous solid cometary nucleus (Whipple, 1950). Assuming that all the absorbed energy is used for sublimation works only well when the comet is close to the Sun: e.g. a water ice nucleus at distances ≤ 1 AU. As was later noted by Crifo (1987), even in the idealized case of a pure solid icy nucleus one should take into account the intermolecular collisions in the sublimating gas. This process results in a significant downward flux of backscattered molecules (up to 25% of the direct flow), that in turn changes the energy balance on the cometary surface and finally, the effective gas production rate.

As the physical models of the nuclear structure become more sophisticated and possibly more realistic (e.g. material may be porous, its composition may include a variety of volatile and non-volatile additives, etc.) determination of the effective gas production is becoming a more and more complicated problem. Various aspects of this fundamental problem of cometary physics were considered by us in numerous articles published over the past few years (e.g., Skorov et al. 2001; Davidsson and Skorov 2002, Davidsson and Skorov 2004). In essence our in-depth analyses revealed the drawbacks of oversimplified thermal models leading to numerous and unpredictable errors in the determination of the effective sublimation rates. Here we focus on another issue that is directly related to the activity problem, but has so far not received due attention. This is the release of gas through a porous non-volatile surface layer of a cometary nucleus. We have dealt with aspects of this problem several times in our previous studies. In the paper Skorov and Rickman (1995) we modeled a kinetic gas flow through a porous cometary mantle. The well-known test particle Monte Carlo method (Bird, 1976) was used. The structure of the porous dust layer was described as a bundle of cylindrical and inclined channels not crossing each other. We investigated how the characteristics of the molecular flow depend on the capillary length, inclination angle, and temperature gradient. The emergent gas flow rate was found to vary with the pore length/radius ratio in excellent agreement with the Clausing formula. Later we presented models (Skorov et. al, 1999; Skorov et al., 2001) simulating the gas emission from ice-dominated walls as well as the outflow through "dusty" channels with a sublimating icy bottom. Furthermore, our models were successfully applied to the interpretion of results obtained in comet simulation (KOSI) experiments (Grün et al., 1993). We also demonstrated that in first order the temperature difference between a sublimating ice front and the surface temperature of an overlying porous dust mantle is almost independent of the thickness of the nonvolatile layer (Skorov et. al, 1999). However, the mere existence of a porous nonvolatile layer on the surface of a cometary nucleus has remained a scientific hypothesis based on the first images of comet Halley's nucleus taken during the flyby of the Giotto spacecraft (Keller et al., 1987). Notice that the low visible albedo of the cometary surface, which is often interpreted as definite evidence of the existence of non-volatile crusts, could just be feigned as demonstrated by a heterogeneous ice-dust model (Davidsson and Skorov, 2002) where a continuous surface crust is not formed. The situation has changed when more recent space missions provided further stunning images of cometary nuclei. To date, we have images of four nuclei of periodic comets. These optical data together with spectral and infrared observations have convincingly shown that cometary nuclei are covered by a porous, dark, non-volatile crust. Exceptionally interesting results were obtained by the Deep Impact mission to comet Tempel I. The thermal inertia of the surface layer was found to be extremely small (Groussin et al., 2007): about three orders of magnitude smaller than the values typical of solid materials. This result clearly indicates the high porosity of the material forming the surface layer. Another observation important for our research was the conclusion that only few traces of water ice were found on the surface. This finding, together with the gas activity measured during the comet Tempel I flyby (Schleicher et al., 2006) can be considered as a strong argument in support of ice sublimation beneath a thin dust layer.

How can cometary activity be understood and modeled under these circumstances? To find plausible answers we assess the models and assumptions used so far in cometary physics to describe the ice sublimation in the porous surface layer. This review follows in the next section. Then, we present a microphysical computational model for molecular transport in a random porous medium and compare the new results with those obtained by similar more idealized models. The focus is on limitations inherent in the considered approaches. In the last section we present our results of the numerical simulation of gas transfer in the near surface region of a cometary nucleus, in an attempt to quantify the various effects.

**Models of gas transfer through a porous dust mantle**

Even a brief overview (Dullien, 1991) of the various approaches and models currently used in the study of gas transport in porous media is an overwhelming task because of the fast development of this field. The situation in cometary physics is quite different and simpler. The majority of publications containing theoretical modeling of gas transfer in the uppermost porous layers of a cometary nucleus uses one and the same basic algebraic formula to calculate effective gas activity - namely, the formula of Knudsen (1950), that describes the mass flow rate per unit capillary area

$$\Psi_K = \left(\frac{32m}{9\pi k}\right)^{1/2} \frac{r}{L} \left(\frac{P_t(T_t)}{\sqrt{T_t}} - \frac{P_b(T_b)}{\sqrt{T_b}}\right) \qquad (2)$$

where $r$ is the channel radius, $L$ is the length, $(P_b, T_b)$ and $(P_t, T_t)$ are pressures and temperatures at the bottom and top of the channel, respectively.

Because this formula is popular in cometary physics and has been used without change for almost thirty years (Fanale and Salvail, 1984; Smoluchowski 1982), we decided to dwell on the physical assumptions for the medium properties used in the derivation of this equation, as well as on the limitations in the application of this formula. The Knudsen formula refers to a very simple model of the porous medium as a bundle of disjoint, straight cylindrical capillary channels of radius $r$, and length $L$ with diffusively scattering walls. The gas in the channel is in the free-molecular regime, i.e. the intermolecular collisions are negligible, whereas scattering by the walls plays a major role. This formula involves the additional supposition that the density gradient is small. The main restriction arises from the assumption that the channels are long and thin ($L \gg r$). This assumption makes its use questionable for a thin porous layer. Thus, all issues related to the formation of the primary dust mantle and outgassing from under a thin layer are not realistically modeled. Therefore rather often a corrective model parameter - tortuosity[*] - is added to the Knudsen formula (Fanale and Salvail, 1984; Huebner et al., 2006). We note that adding of new model parameter does not extend the applicability of the Knudsen formula.

Fortunately there exists a simple generalization of the Knudsen approach. It allows us to calculate the rarefied gas flow in a Knudsen regime through a cylindrical tube of arbitrary length with diffusion scattering walls with high accuracy. This is the

---

[*] There are various mathematical methods of tortuosity estimation in 2D and in 3D. The simplest way is the ratio of the actual path through the porous layer to its thickness (so called arc-chord ratio).

Clausing formula which apparently was for the first time considered in cometary physics by Steiner (1990):

$$\Psi_C = \left(\frac{m}{2\pi k}\right)^{1/2} \frac{20+8(L/r)}{20+19(L/r)+3(L/r)^2} \left(\frac{P_t}{\sqrt{T_t}} - \frac{P_b}{\sqrt{T_b}}\right) \quad (3)$$

He also performed a quantitative comparison of the Knudsen and the Clausing formulas and showed that even for a sufficiently long channel in which the ratio of length to radius equals 10, the gas flow calculated using the former formula is overestimated by 50%, while for short tubes ($L \approx r$) the relative error is about eight times higher. We emphasize again that the Clausing technique yields the exact solution. Later Skorov et al. (1999, 2001) applied this approach for modeling a gas flow through a cylindrical tube with icy walls of varying temperature. Obviously, the main difficulty derives from the fact that molecules do not only fly through a tube, but that some are born on the icy walls as a result of ice sublimation. Unfortunately neither the paper of Steiner (1990), nor the later work of Skorov and Rickman (1995) have attracted appropriate attention from comet researchers. It is all the more incomprehensible, because like the Knudsen formula the accurate Clausing expression has also a simple algebraic form and uses the same set of model parameters.

If the Clausing formula is exact and valid (Clausing 1932) for channels of arbitrary length, then what is the difficulty of modeling the transport of sublimation products? The answer is simple: the difficulty of modeling lies in the appropriate description of the properties of the porous medium. Both formulae considered above were obtained for the molecular gas flow in a tube, while our ultimate goal is the calculation of gas flow in natural stochastic porous media, where statistically the length of a void is the same in any arbitrary direction. The transition from the model of a single pipe to the model of porous medium can not be realized in a simple way when the capillary approach is used. The extension of the results obtained for cylindrical tubes to the case of a stochastic medium is a non-trivial task. For example, in a capillary model the porous medium is simulated as a bundle of parallel tubes, and therefore the macroscopic volume porosity of the model medium is determined as $\pi r^2 N$, where $N$ is the number of tubes per unit area. It is apparent that this model medium is clearly anisotropic because the porosity is different in different directions and/or for different control surfaces. Moreover, for an arbitrary number of tubes per unit area (hence for arbitrary volume porosity) one can always construct a control surface, which does not cross the capillaries, and therefore has zero surface porosity. Thus, the anisotropic character of the capillary model medium is a serious obstacle for its use in cometary physics applications. Nevertheless till now the majority of cometary models are built on the capillary porosity model.

**Tortuosity**

In order to compensate for this problem an additional model parameter the above-mentioned tortuosity, τ, is usually added. The formal purpose is to replace a straight cylindrical channel by a broken one of greater length and thereby to make the model environment more isotropic. This idea is rather popular with cometary researchers and used by many authors. However, different authors use different values for this parameter: in (Kossacki and Szutowicz, 2008; Mekler et al., 1990) the tortuosity is equal to $\sqrt{2}$, in (Prialnik and Podolak, 1999; Whipple and Stefanic, 1966) the tortuosity is equal to 4.6, in (Fanale and Salvail, 1984) the tortuosity is equal to 5, in Enzian et al. (1997) the tortuosity is equal to 13/9 ($\approx \sqrt{2}$), in Huebner et al. (2006) the tortuosity is treated as a free model parameter varying from 1 to 2.

The reason for these discrepancies derives from the fact that the tortuosity is introduced in a non-unique way into the simplest capillary model where the gas travels through the media following a winding path approximated by bundles of parallel straight cylindrical tubes. For a natural porous media the tortuosity is an empirical quantity, that should be determined from the experiments. Note that, in contrast to porosity, tortuosity can not be easily measured directly. Often it is derived from independent measurements of the Knudsen diffusion coefficient and porosity. In this case the tortuosity is retrieved from the simple expression connecting diffusion coefficients in a free phase $D_0$ and in a porous medium $D_p$

$$D_p = \frac{\varphi}{\tau} D_0, \tag{4}$$

where τ is the tortuosity factor (Carman, 1956). This formula shows that the presence of solid scattering obstacles causes the diffusion paths to deviate from straight lines. As a result, the effective diffusion must be corrected for the tortuosity.

Tortuosity is defined in various ways in scientific literature. This often leads to misunderstanding, because no consensus exists how it should be determined. Tortuosity is generally defined as the ratio of effective path (or actual length $L_e$) to thickness of the porous layer $L$, or as the ratio of the average free path traveled by the species $\langle l \rangle$ to the average displacement in the direction of diffusion $\langle x \rangle$ (e.g. Boudreau 1996)

$$\theta = \frac{L_e}{L} = \frac{\langle l \rangle}{\langle x \rangle} \qquad (5)$$

It is important to note, that the quantity "tortuosity factor" is also widely used in the literature in parallel with tortuosity. The former is usually included in Fick's law and equal, following Carman (1937), to $\theta^2$. However, this definition is not generally accepted, and some authors define "tortuosity" instead of "tortuosity factor" as $(L_e/L)^2$ (e.g. Dullien 1991, Salem 2000).

There are many theoretical models that calculate the tortuosity for a specific structure of a porous medium. The simplest theoretical models do not include an adjustable model parameter and are rather idealized. It is interesting to note that even in the simplest case a different treatment of the capillary structure leads to different results for formally identical porous media. For example, Bhatia (1996) presents a transport model under the assumption of uniformly sized cylindrical pores and random pore orientation. For this case the tortuosity factor is equal to $1/<cos^2\gamma>$ (Johnson and Stewart, 1965), where $\gamma$ is the angle between the net diffusion direction in a pore and the macroscopic flux vector. With an additional assumption that diffusion in a pore proceeds axially the tortuosity is then equal to $\sqrt{3}$. Markin (1965) considers a similar model, but writes that the mean free path of test particles $\sim \varphi/(1-\varphi)$, leading to a tortuosity equal to 2. Petersen (1958) uses $\theta = \sqrt{2}$. Measured values of the tortuosity often correlate with porosity. This leads to more complicated models where the tortuosity depends on porosity $\varphi$. The seed of this idea goes back to papers of Maxwell (1881) and Rayleigh (1892). We do not consider here these models and refer the reader to Table 1 in (Boudreau 1996), where some expressions are collected. We only note that for most models as porosity approaches unity tortuosity approaches unity too. This means there is no hindrance for diffusion in the absence of scattering solids.

**Models of packed beds**

So far we considered models where the pores are described as a system of capillaries (vertical, inclined or twisting). Usually it is assumed that pores are isolated from each other, rarely more complicated models are used (Bhatia, 1996), where pores can be closed and/or cross each other. In all cases, the gas flow through channels, i.e. the internal molecular flow through a solid porous medium with a high level of connectedness, is investigated. We noted above the complexity involved in applying this approach to the description of the transport of sublimation products through a porous surface layer of a cometary nucleus. There is an alternative way to describe the Knudsen diffusion in a porous medium - a way where the molecular gas flow is regarded to be external to the nonvolatile matrix: "flow around obstacles". The matrix itself is constructed (composed) of elementary scattering or absorbing objects, for example, spheres. Interaction of gas molecules with the surfaces of matrix elements is similar to interaction with the walls of a capillary in the models of the first type. Thus, the weakening of gas flow and reduction of effective diffusion rate can be well modeled as before.

This description of the structure and porosity of the surface layer of cometary nuclei seems more natural and attractive. From the huge amount of photometric and polarimetric observations of cometary tails (Fulle, 2004) one can conclude that dust tails are preferentially formed by particles that are likely to have appreciable porosity and a complex structure. Body resolved observations yield very low values for the thermal inertia of the surface layers of cometary nuclei (Groussin et al., 2007; Groussin et al., 2010) suggesting that they consist of very loose, porous, and weakly bound dust. This, apparently, explains the observed high surface temperatures and absence of large surface areas of ice inspite of significant observed gas release. Conclusions drawn from these observations are consistent with results of theoretical models that address the removal of nonvolatile material by gas flow while a thin non-volatile layer is maintained on the nucleus surface (Skorov et al., 1999; Huebner et al., 2006).

Thus we arrive at the so-called model of a granular packed bed that is widely used e.g. in catalysis and powder researches (Wen and Ding, 2006; van der Hoef et al., 2006), but practically not applied in cometary physics. An exception is Davidsson and Skorov (2002), where gas transport through a random porous surface layer via Monte Carlo simulation was investigated. The packing consists of elementary objects (e.g., spheres or cubes) randomly or regularly distributed in many ways. The properties of the resulting layer (porosity, permeability) can vary in a wide range. Hereafter we treat permeability as a measure of the ability of a porous medium to transmit gases. The minimal porosity in a monodisperse regular packed bed (hexagonal close packed lattice) is about 26%, in a random monodisperse layer the minimal porosity is about 36,6% (Song et al., 2008), whereas in a random medium constructed by gravitational deposition the porosity can be as high as 85% (Blum and Schräpler, 2004). The next refinement of the model of the surface layer can be made based on the analysis of the formative processes.

Assuming that the dusty layer observed on the nucleus surface is the result of ice sublimation, and as a consequence, of deposition of initially porous dust-ice material that lost its volatile component one can make the following model assumption. During the layer formation, dust particles actively interact with a gas flow directed normal to the surface, so the spatial structure of the residual layer should be similar to the structure of a porous medium formed by gravitational sedimentation. This model assumption immediately gives us the expected range of porosity: from 65% to 85%.

**Porous media generation**

We generate porous media consisting of monodisperse spheres using two different methods: ballistic deposition (RBD) and random sequential packing (RSP). All media are cubical with sides of 50 sphere diameters ($d$). For RBD spheres are dropped one by one vertically into a control volume where they touch either the bottom or another sphere. The process is terminated when a sphere reaches the top of the control volume, or when a predetermined number of spheres is reached. Visual inspection of the medium reveals branching and filamentous structures preferentially in vertical direction. The porosity of media generated by RBD is about 0.85, with a more compact base and a fluffier top. In order to avoid boundary effects and to produce a medium with an approximately homogeneous porosity, we cut out the central part (70$d$ x 70$d$ wide and 100$d$ high) of a larger medium that is used in a random walk model. For RSP spheres are placed one by one at random locations within a control volume. Locations that would result in overlapping spheres are rejected. This results in a homogeneous, isotropic medium with a porosity that can be determined by specifying the total number of spheres used. The lowest porosity that can be reached using this method is about 0.63 (Torquato et al., 2006). We generated media with porosities of 0.65, 0.70, 0.75, 0.80 and 0.85. In both cases there are spatial variations of porosity over the column cross section in the sample. These variations depend on the sample structure. For the RPS sample variation of porosity is small and has the same amplitude for different cross sections. For the RBD one can expect that the porosity may be slightly different in the vertical and in horizontal directions due to the initial anisotropy of the generation process.

**Gas transport through porous media**

In order to estimate the transport properties of the model media we apply the random walk algorithm used extensively in studies of disordered granular media (Levitz, 1997). The major characteristics of a porous layer that are important for cometary applications are: i) the permeability that determines the effective gas production from under the porous layer; ii) the resulting return flux that determines the re-condensation rate and, in turn, the effective energy loss due to sublimation, and iii) the diffusion rate through the porous dust layer that determines the characteristic time scale for the gas diffusion processes (e.g., for gas pressure relaxation). These characteristics are examined by tracing the geometric paths of a large number of test particles through the medium using the Monte Carlo method. We use 100.000 test particles for each simulation. Each particle starts at the bottom of the medium in a random upward direction. Its path through the medium is traced until it exits from the top or the bottom of the medium. Particles exiting from the sides of the medium are accounted for by reinserting them from the other side. Since the outflow is assumed to be rarefied (the thermodynamic molecular free path is much larger than a sphere size), intermolecular collisions and particle velocities are not considered. Therefore, the geometric paths of test particles only are a result of their starting direction and subsequent interactions with the spheres that represent the nonvolatile matrix. These interactions are modeled as either specular reflections (incoming angle equals outgoing angle) or diffuse scattering (the test particle is redirected in a random direction away from the sphere). During the simulation we collect information about the free path in the media in order to calculate the mean free path distribution and average vertical displacement. Finally, the maximum height reached by each test particle in the medium $h_m$ is recorded. Then the permeability $\vartheta(h)$ is found as a function of height $h$ by:

$$\vartheta(h) = \frac{Number\_of\_particles(h_m \geq h)}{Total\_number\_of\_particles} \tag{6}$$

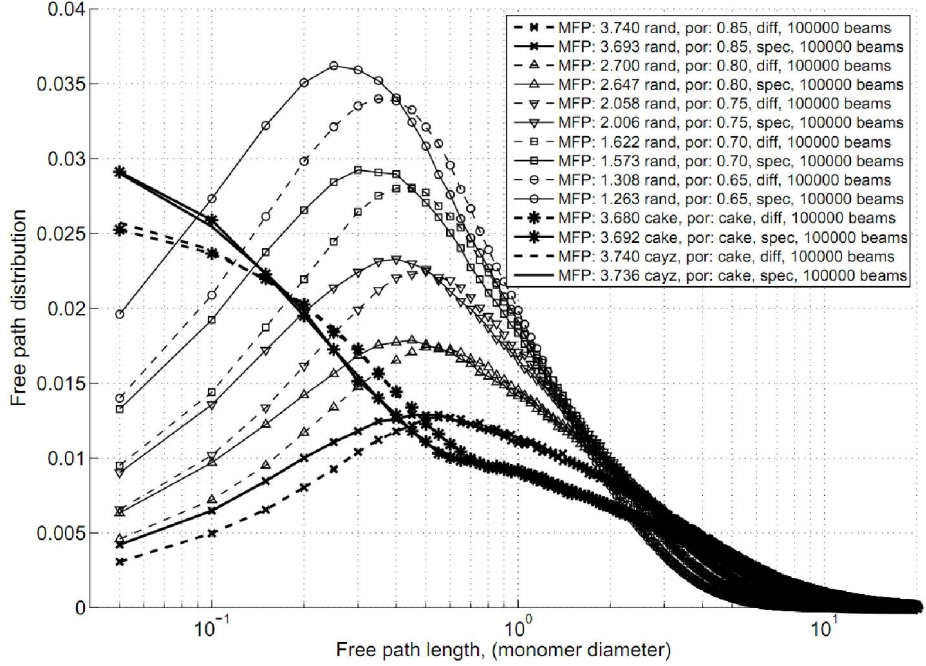

Figure 1. Distribution of the mean free path for different models of porous media: the RBD model and the RSP models with diffuse and specular scattering.

**Results**

At first we examine the geometry of the resulting porous media, looking at the distribution of the free path (FPD) of the test particles in the samples. This distribution (or, what is the same, the chord length distribution) is connected to the effective pore size and, hence, to the Knudsen diffusivity (Levitz 1993; Levitz, 1997). The obtained results for RBD and RSP are shown in Fig. 1. We present the FPD both for media with diffuse and specular scattering. In all considered cases the scattering law is of no consequence for the high porosity range considered. The distributions become identical as the sample thickness increases. The comparison of results obtained for consolidated (RBD) and partially unconsolidated (RSP) media is more interesting. We see that in the RBD cake the short chords play the major role. The dominant contribution to the FPD for small chords comes from contacting pairs of spheres (Pavlovitch et al., 1991). The FPD for the RBD model fits an exponential form well (Dixmier, 1978; Pavlovitch et al., 1991). For the RSP model the contacts between spheres are rare. This leads to the quite different width of the FPD at the range of small chords.

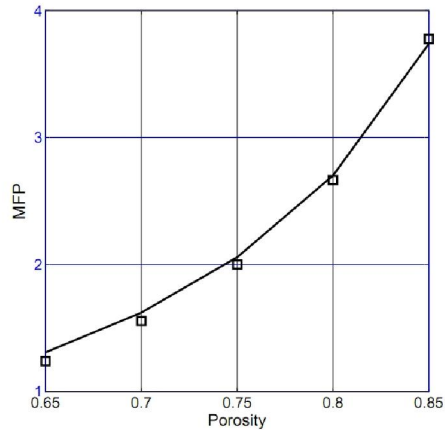

Figure 2. Mean free path for the RSP model as a function of porosity: simulation (squares) vs. theory (solid curve).

If we investigate the curves obtained for the RBD and RSP models with the same porosity ($\varphi=0.85$) we see that the difference between the FPDs is visible for all the chord lengths. The position of the maximum of the FPD does not depend on the sample porosity, whereas its value is sensitive to the filling factor (i.e. to the effective number density of the sample). The calculated distributions clearly show that the chord length of void space can be much larger than the sphere size. This feature is worth noting because it means that both mass and energy balance in an elementary volume of a highly porous cometary surface layer are determined not only by local values of the thermophysical characteristics, but also by their values in the surrounding region, the size of which far exceeds the size of the dust particles and/or pores. It is interesting to compare values of the mean free path (MFP) for different model media. As one might expect the MFP decreases rapidly when we reduce porosity (Fig. 2). Thus the decrease of the filling factor when the porosity changes by 30% (from 0.65 to 0.85) leads to an increase of the MFP approximately by a factor 3. The obtained dependence is in good agreement with theoretical predictions (Dullien 1991): $MFP \sim \varphi/(1-\varphi)$. We will return to this characteristic later when we discuss the tortuosity and the correlation of our models with the idealized Clausing capillary model.

Now we proceed to the examination of transport properties of the investigated porous media: namely the permeability and the estimation of the return molecular flux.

As a first step we investigate the role of the scattering law. The normalized permeability for the random media with specular and diffuse scattering is shown in Fig. 3. For a fixed porosity the permeability is insensitive to the law of scattering. From here on, without loss of generality, we consider only diffuse scattering type.

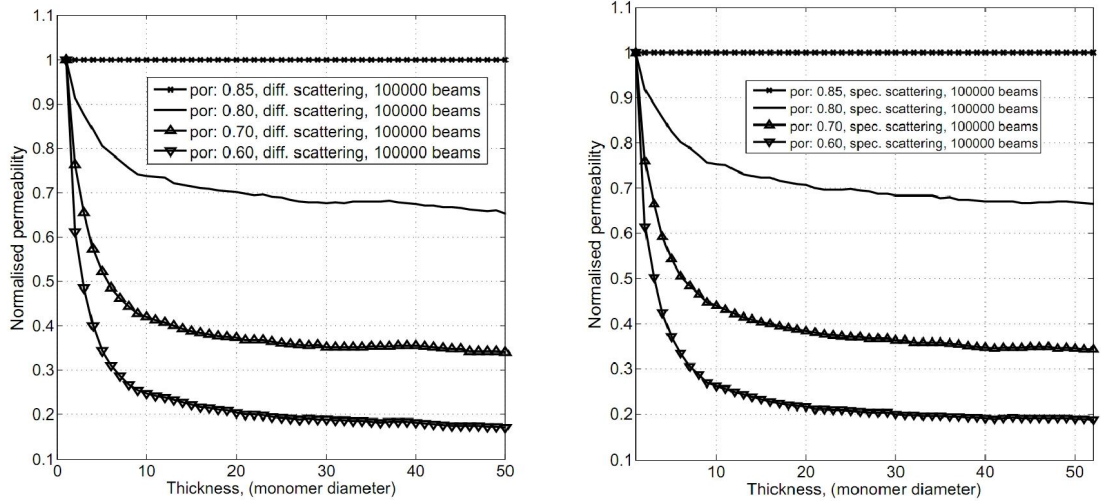

Figure 3. Normalized permeability as a function of layer thickness for the RSP models with diffuse (panel a) and specular (panel b) scattering.

As next step we investigate how important the spatial organization of scattering spheres is. Two models with the same porosity ($\varphi=0.85$) are tested with the RBD cake, having inherent anisotropy, and the RSP sample. The mean values of the free path are similar whereas the FPDs are of similar character only for long chords (see Fig. 1). The normalized permeability of the RBD cake in two orthogonal directions is plotted in Fig. 4. As might be expected the permeability is different in the vertical and horizontal directions, and the deviation from the permeability of isotropic random media grows when the anisotropy level of the sample is increased (see panel b). The RBD model seems physically more reasonable, but cannot be characterized by a mean porosity only. It has anisotropic transport characteristics. Therefore we recommend to use this model with care.

The isotropy of the RSP model for different values of porosity is presented in Fig. 5. The vertical permeability of the random medium with $\varphi=0.85$ is used as normalizing function. The variation of permeability in different directions becomes more visible for the compact case ($\varphi=0.65$). But even in this case the difference is relatively small and unimportant for our investigations. We consider the RSP model to be isotropic for the whole range of porosities.

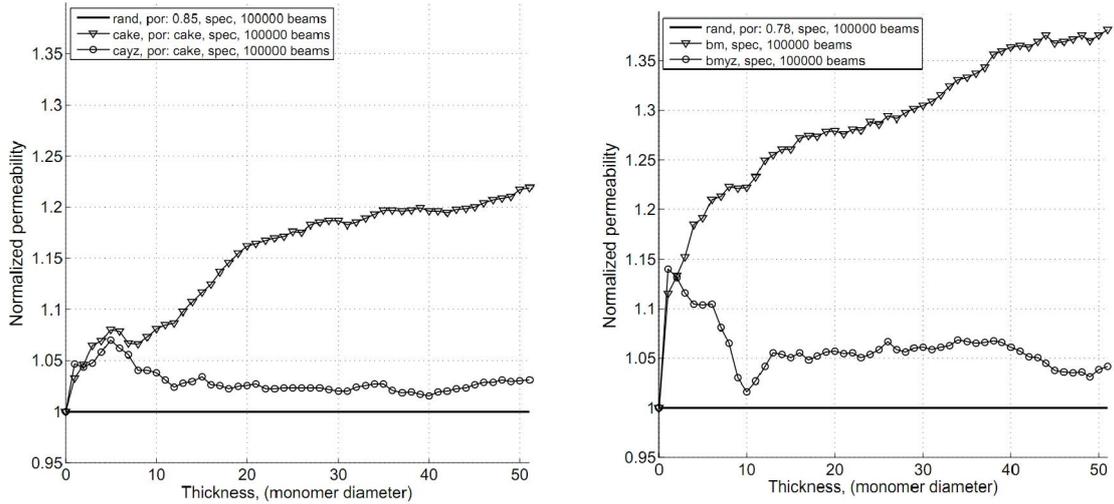

Figure 4. Normalized permeability as a function of thickness for the RSP model and the RBD model having the same average porosity. Porosity equals 0.85 (panel a) and 0.78 (panel b).

Just as the MFP drops down when the porosity is decreased, the permeability also diminishes rapidly. The obvious question arises about the functional connection between permeability and MFP for isotropic random loose media. The normalized permeability as a function of porosity is shown in Fig. 6 for three different thicknesses of the porous layer. The dependence for the MFP obtained earlier is also plotted in the normalized form. One can see that at least for thick (extended) layers the functional behaviors are similar and hence we can conclude that the MFP can be used to describe the permeability of loose RSP media.

For cometary applications an important characteristic of the medium is the flux of scattered molecules returning to the sublimating surface (i.e. the number of re-condensed molecules). It is this return flux that influences the effective sublimation rate and hence the overall energy loss due to sublimation. Because sublimation plays a major role at small heliocentric distances its accurate calculation is of practical importance for cometary activity in the inner solar system. Results of the simulation are shown in Fig. 7, where the normalized return fluxes are plotted as a function of layer thickness for different sample porosities. As before the result obtained for the random medium with $\varphi=0.85$ is used as a normalizing function. The relatively small variation of porosity (about 30%) leads to substantial changes of the return flux. These changes are still significant for a layer as thick as 20 sphere radii. The variation of the return flux arising from structural organization of the scatterers (with the same porosity $\varphi=0.85$) does not exceed 10% and practically vanishes when the layer thickness is larger than 20 radii. Thus we conclude that both the variation of porosity and the structure of the layer are of importance for the gas flux leaving the medium.

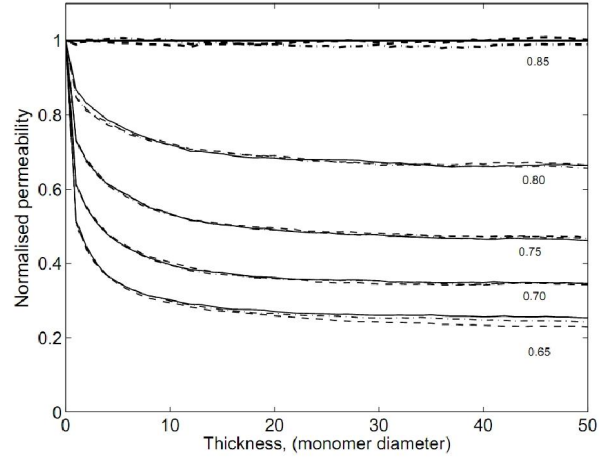

Figure 5. Normalized permeability of the RSP models with specular scattering for different values of porosity. The vertical permeability of the random medium with $\varphi=0.85$ is used as normalizing function. Permeability is calculated in vertical (solid curves) and two horizontal (dashed and dashed-dotted curves) directions. The corresponding porosities are also shown.

These results allow us to reconsider the model of capillary transport of gas discussed above. Despite the numerous shortcomings of this model, it has one indisputable advantage that explains its popularity among cometary researchers. Its virtue is the simplicity of the formulas (Knudsen and Clausing) describing the gas flow in a porous medium, and hence the ease with which the description of this process can be included in the computational model of heat and mass transfer. When a typical problem of a transient temperature distribution in the surface layer of a cometary nucleus is solved the gas flow as a function of the current temperature distribution and spatial coordinates has to be recalculated at each time step. Unfortunately, any accurate statistic simulation requires a large number of test events. This makes it impracticable to apply the method of statistical modeling directly as it is used to simulate transport through a random porous medium in the frame of a general thermophysical model. We see two possibilities to circumvent this problem. The first possibility was used by Davidsson and Skorov (2002). For the one-dimensional model of heat and mass transfer in a porous ice layer we carried out many calculations of the sublimated gas flow in advance for the entire range of expected values of temperature and its gradient. These stored arrays of numerical data were later used to calculate the temperature field. The second possibility is to build an empirical algebraic formula similar, for example, to Clausing's formula. The parameters of this expression should be based on (or retrieved from) more realistic and complex models of the porous medium such as presented above. The results presented in Fig. 6 show that the relationship between permeability (and with it effective sublimation) and porosity of the medium is not linear, whereas the terms for the Knudsen as well as for the Clausing formula are linear.

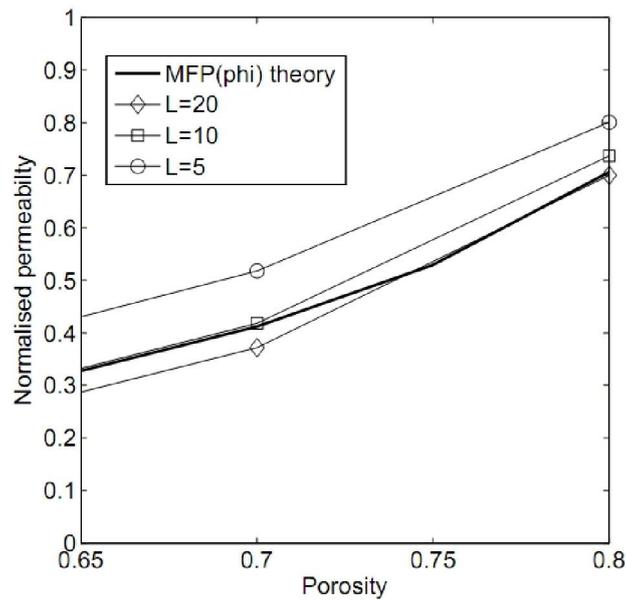

Figure 6. Normalized permeability as a function of porosity for the RSP model is plotted for layers of different thickness. The calculated mean free path is presented for comparison.

A comparison of the resulting gas flow calculated using the classical Clausing formula with the gas flow obtained from RBD and RSP models is shown in Fig. 8. For layers with a porosity of $\varphi=0.85$ the difference in the flow can reach 20% if the thickness is smaller than 100 radii. This difference is observed for all tested models both with diffuse and specular scattering. Obviously, tortuosity can not be entered into the required algebraic expression as a simple factor. Thus, one needs to find a more complex relationship without increasing the number of model parameters. As a possible solution we propose: i) to retain the Clausing formula since it accurately describes the flow of gas through a channel of arbitrary length; ii) to not add "external" factors associated with macroscopic characteristics of the environment such as porosity and tortuosity; iii) but to introduce instead "an effective radius" of the capillary, that must be calculated based on statistical modeling. Minimizing the difference between the results obtained for capillary and RSP models we find the corresponding value of the effective radius $r_{eff}$ and the ratio $r_c=r_{eff}/r$ for each given value of porosity. The latter is shown in Fig. 9. It differs from the dependence retrieved earlier for the MFP. Hence the MFP cannot to be used as a substitute for "an effective radius" of a capillary. The value of $r_c$ as a function of porosity is well approximated by the formula: $r_c = 17.4\varphi^2 - 22.3\varphi + 7.4$. This simple approximation holds to within 4%.

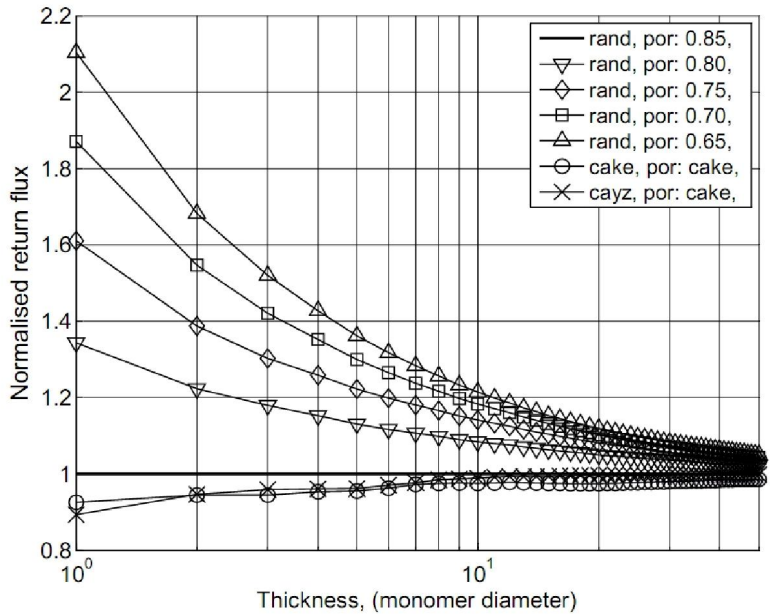

Figure 7. Normalized return flux as a function of the layer thickness for the RSP models with different porosity and for the RBD model. Specular scattering was used for the all cases.

**Summary and Discussion**

Accurate determination of the effective sublimation rate or, in other words, of the net mass flux from a cometary nucleus is one of the actual problems of the cometary physics. *In situ* space observations provide strong arguments that cometary nuclei are largely covered by porous nonvolatile material. Thus the gas release depends on the details of the kinetics of molecule transport through this permeable layer.

The present paper aims at three primary goals:

1. Revise or adjust the capillary models used in cometary physics to describe the transport of sublimation products through porous nonvolatile layer accurately.

2. Present alternative description of porous media based on ballistic deposition and random sequential packing methods. Use direct statistical simulation to retrieve major geometrical and transport properties of model media as a function of porosity and layer thickness.

3. Suggest a way to adjust the Clausing formula taking into consideration the nonlinear dependence of permeability on layer porosity.

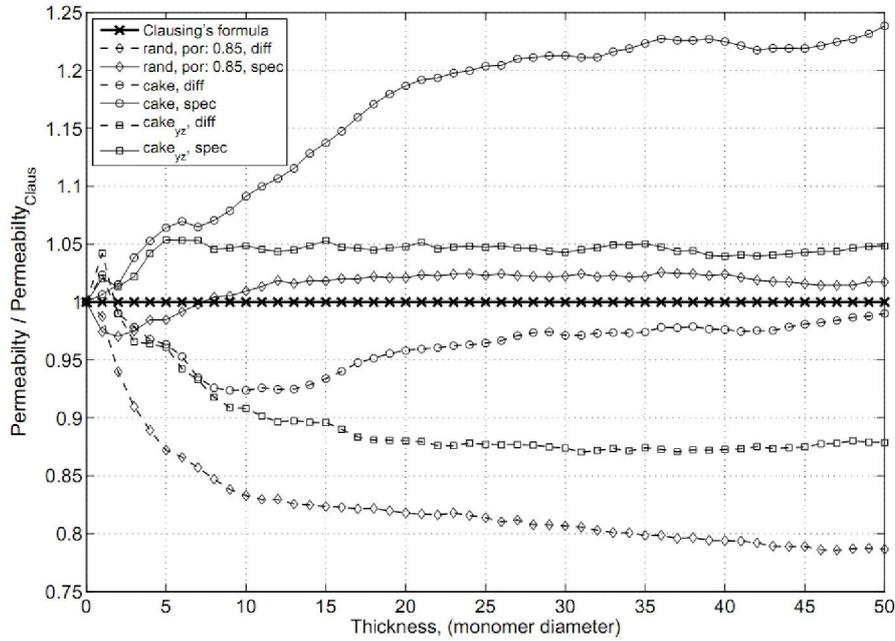

Figure 8. Comparison of the permeability calculated for the different models of random porous media ($\varphi$=0.85) in different directions with the permeability calculated by the fitted Clausing formula (used as a normalizing function). Diffuse and specular scatterings are presented for the random porous models.

In view of these goals we summarize below the main results:

1. Knudsen's formula widely used in cometary research is not applicable for modeling the gas transport through short channels. Satisfactory agreement with experimental data is achieved only when the channel radius is much smaller than its length. This fundamental shortcoming is not eliminated by introducing additional phenomenological parameters (e.g. tortuosity). Hence, the effect of a porous dust mantle on the surface for the sublimation rate of a cometary nucleus cannot be accurately simulated using this formula. This fact was previously pointed out by Steiner (1990) and Skorov and Rickman (1995). As an alternative, Clausing's formula can be used, which is still rarely incorporated into theoretical models describing transport processes in the surface layers of the nucleus. This formula gives an exact agreement with the experimental data for straight cylindrical channels with an arbitrary ratio of radius to length. However, the approach using the Clausing formula has still shortcomings, characteristic for models where the natural porous medium is described as a bundle of straight cylindrical capillaries. The spatial anisotropy of the capillary model leads to the fact that its transfer characteristics are highly different for different directions. The transition from the permeability of one channel to the permeability of the medium can not be accurately and correctly generalized. Adding an additional linear factor - tortuosity to the formula does not solve the problem, but on the contrary, only confuses the situation. This parameter depends on the porosity of the medium, and its value cannot be uniquely determined for Knudsen diffusion through the pores, when the collisions of molecules with the scattering dust fraction play a dominant role.

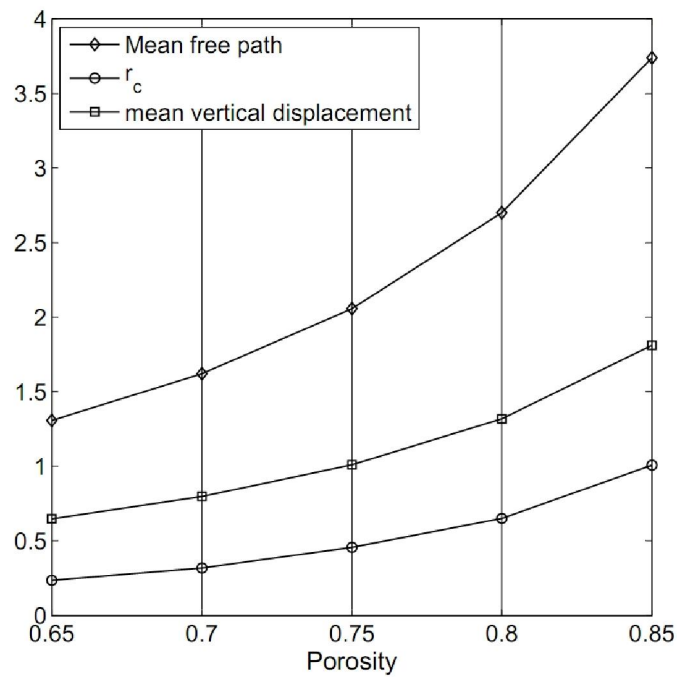

Figure 9. Dimensionless ratio $r_c$ used in the fitted Clausing formula, mean free path and mean vertical displacement as a function of porosity are presented for the RSP models with specular scattering.

2. To avoid these inconsistencies accurate statistical calculations are performed for media formed either by ballistic deposition (RDB) of test particles or by random filling of a control volume (RSP). To determine the degree of anisotropy of the constructed media the average porosity is analyzed in different directions. It is shown that the RSP model has a high degree of isotropy, whereas the medium constructed by the RBD procedure has an inherent anisotropy. We calculate the distribution function of the free path (FPD) in model media with different porosity. In the RSP model neighboring scattering spheres are not always in contact with each other. This significantly alters the FPD, but the mean free path changes very little and is mainly a function of the medium porosity. The calculated nonlinear dependence of the mean free path on the porosity is in agreement with its expected theoretical dependence. At the next step the medium permeability and the return flow of re-condensed molecules are estimated for the different model parameters. Two types of interaction of molecules with scattering spheres are tested: diffuse and specular scattering. It turns out that for the random model of the porous medium the permeability is virtually independent of the type of interaction. This feature, of course, differs fundamentally from the properties of the capillary model, where the permeability of the medium in the case of specular scattering is always unity irrespective of the capillary geometry. We show that a relatively small variation of porosity (not more than 30%) leads to a strong change of permeability. The permeability depends on the medium porosity in a nonlinear way. Thus, we conclude that capillary models of porous media commonly used in cometary physics are of limited accuracy. At the same time, our extended calculations show again that direct statistical modeling requires a lot of computer time and power. This makes it impracticable to directly include statistical computations in a general model of heat - mass transfer in the surface layers of a cometary nucleus.

3. In order to overcome the resource consuming calculations for direct use of the statistical models, the approach used by Davidsson and Skorov (2004) could be applied. In this paper we have presented a different way to calculate the effective permeability. We preserve the overall structure of Clausing's formula, accurately describing the kinetics of transport through a single cylindrical capillary. To take into consideration the porosity of the medium in an appropriate manner we assume that the effective radius of the capillary is an unknown function of porosity. The explicit form of this functional dependence is derived from a nonlinear approximation based on statistical modeling results. Thus, the effective permeability, as before, depends on the thickness of the layer and its effective pore size, which in turn is a function of porosity. The retrieved algebraic expression allows us to accurately calculate the permeability of layers whose thickness and porosity vary in the range of values expected for the near-surface regions of a cometary nucleus. The simplicity of this approach makes it practical to include the computational block that accurately describes the transport of gas in the overall thermal model of a cometary nucleus.

This work was supported by the German Research Foundation (DFG grant BI 298/9-1).